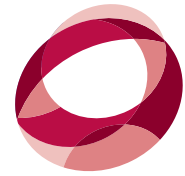

# 5G Security and Privacy – A Research Roadmap

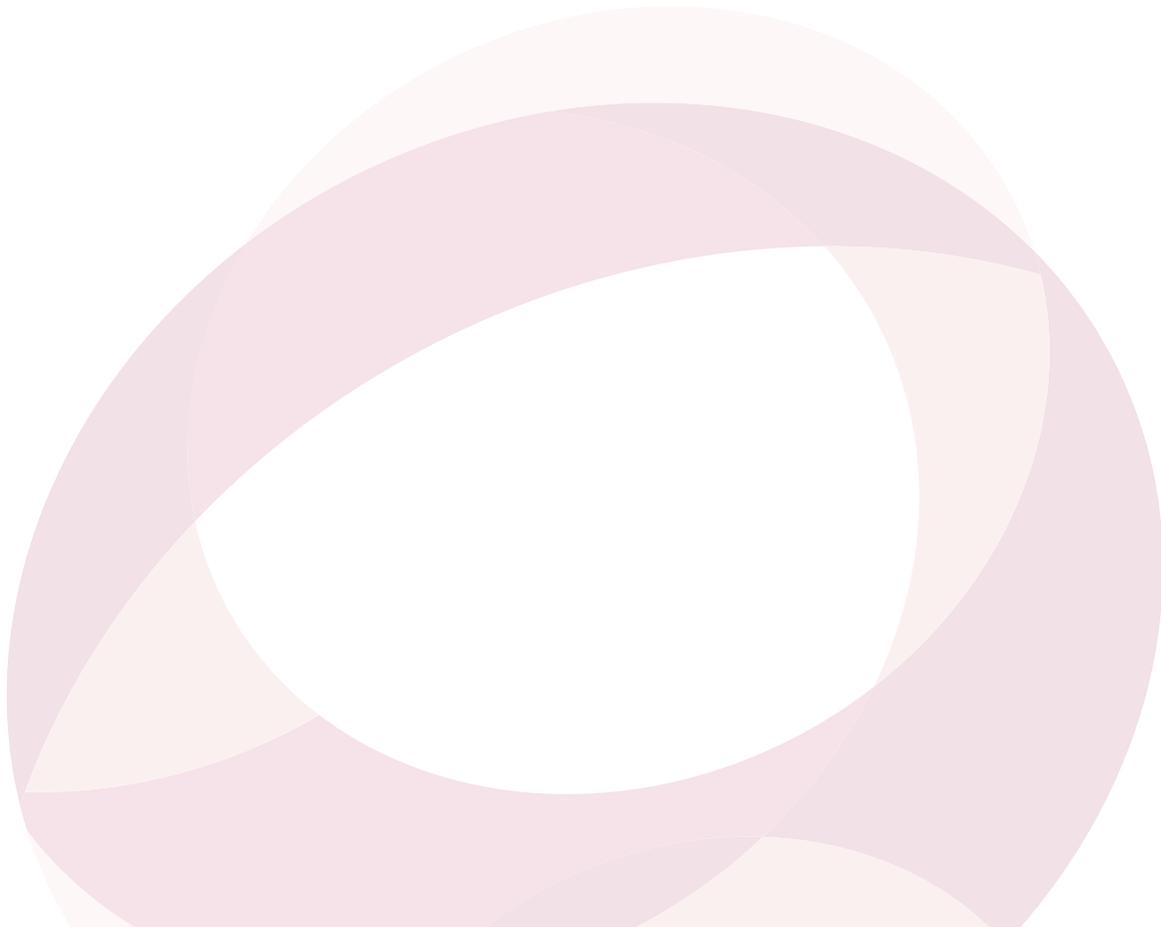

Elisa Bertino and Syed Rafiul Hussain
CS Department, Purdue University, {bertino, hussain1}@purdue.edu

Omar Chowdhury, CS Department, University of Iowa, omar-chowdhury@uiowa.edu

*Abstract:* Cellular networks represent a critical infrastructure and their security is thus crucial. 5G – the latest generation of cellular networks – combines different technologies to increase capacity, reduce latency, and save energy. Due to its complexity and scale, however, ensuring its security is extremely challenging. In this white paper, we outline recent approaches supporting systematic analyses of 4G LTE and 5G protocols and their related defenses and introduce an initial security and privacy roadmap, covering different research challenges, including formal and comprehensive analyses of cellular protocols as defined by the standardization groups, verification of the software implementing the protocols, the design of robust defenses, and application and device security.

This material is based upon work supported by the National Science Foundation under Grant No. 1734706. Any opinions, findings, and conclusions or recommendations expressed in this material are those of the authors and do not necessarily reflect the views of the National Science Foundation.

# 1. Introduction

The Fourth Generation Long Term Evolution (4G LTE) technology has increased the bandwidth available for smartphones, in essence, delivering broadband capacity to smartphones. The most recent 5G technology is further enhancing the transmission capacity and reducing latency through the use of different technologies. It is expected to provide Internet connections that are at least 40 times faster than 4G LTE.

Cellular networks are undoubtedly one of the most critical infrastructures. The novel 5G cellular networks [21] will connect IoT devices and systems, and thus promises to contribute to the transformation of cities, homes, healthcare imaging and diagnostics, manufacturing, transportation, and robotics.

However, because 5G promises to make cellular networks pervasive and able to be used in any application we may think of, ensuring the security of 5G cellular network is highly critical [20]. For example, a denial of service (DoS) attack carried out on a 5G cellular network may paralyze entire communities and service infrastructures with disastrous consequences.

Securing 5G cellular networks is a challenging and involved task. The 5G network protocol stack consists of multiple layers, e.g., physical layer, radio resource control (RRC) layer, non-access stratum (NAS) layer, etc. Each layer in turn has its own protocols to implement its procedures, such as the protocols for connecting/disconnecting devices to/from the network and for paging devices to deliver notifications of incoming calls and SMS. In addition, vulnerabilities can be introduced in the implementation of those protocols as the protocols are complex and the standards include many different options and leave some requirements abstract and implementation dependent. Additional requirements, such as backward compatibility, add to this complexity. It is important to point out that even though 5G is similar to 4G in all those respects, 5G will have additional challenges due to the many novel technologies that are (or will be) incorporated into it (see Section 2), which in turn will require adding new protocols and/or changing existing ones. Many such technologies have not yet been analyzed with respect to security.

Further away from previous generations, 5G network has introduced major changes in the protocol stack and system architectures. For instance, 5G physical layer supports mobile broadband, massive machine type communication, and ultra-reliable and low latency communication for a wide array of devices and applications. It also supports the public-key encryption in the non-access stratum (NAS) layer, concurrent multiple-registrations for connection through heterogeneous radio access technologies (e.g., 5G RAN and WLAN), different types of handover to support mobility, and service-based architecture at the core network that is required to incorporate new technologies and security mechanisms in the protocol, such as network slicing, programmable networking, and protections against unauthenticated access to network slices. As a result, along with the vulnerabilities in the inherited functionalities from 3G/4G networks, the new technologies and protocols added in 5G networks introduce new attack surfaces that have not yet been analyzed with respect to security and user privacy. So even though 5G has introduced security improvements over 4G, new vulnerabilities are likely to have been introduced because of the major extensions that 5G has introduced over 4G.

Comprehensive analysis of such complex networks is challenging because of the complexity of the protocols, e.g., there are multiple participants and many intertwined sub-protocols across multiple protocol layers where each layer has its own protocols. In addition, reasoning about cryptographic constructs adds more complexity to the analysis.

Research in the past few years has succeeded in identifying vulnerabilities that not only jeopardize the security of the ecosystem but also impact user privacy. However, this is just the tip of the iceberg and we are far from having systematic and comprehensive approaches to identify such vulnerabilities. In some cases prior work has relied on the analysis of network traces, in other cases on the use of cryptographic protocol verifiers, and in many other cases on researchers' intuition. As a result, prior analyses are limited in scope and effectiveness. Devising suitable defenses against identified vulnerabilities is equally critical. However, deploying and/or extending available security techniques for wide-spread use in cellular networks is also challenging due to the complexity of the cellular network ecosystems and the incentives of the different stakeholders.

The goal of this white paper is to propose an initial research roadmap, which can be augmented by interested parties, in order to open discussions around the security and privacy of 5G technology.

## 2. 5G Technology

5G cellular networks are based on a number of different technologies, which we review below:

- **Millimeter waves:** These are electromagnetic waves that lie in the frequency range of 30-300 GHz in contrast to the band below 6GHz used for 4G LTE [1]. The microwave band is just below the millimeter-wave band and is typically defined to cover the 3–30 GHz range. Millimeter waves allow 5G networks to transmit very large amounts of data but only at short distances, and to also use unlicensed frequencies, currently used by Wi-Fi, without creating conflicts with Wi-Fi networks through the use of small cells to complement conventional cellular networks.

- **Small cells:** These cells are portable base stations that complement conventional macro cells by providing extended coverage and increasing network capacities on demand. In addition, due to the reduced transmitter-receiver distance, small cells enhance energy efficiency [2]. Their intended use is mainly for dense areas, such as stadiums, and indoors. As a result 5G networks will rely on a multifaceted infrastructure consisting of macro and small ultra-dense cells. The fact that small cells are portable will enable dynamic infrastructure.

- **Massive MIMO:** Massive multiple-input, multiple-output, or *massive MIMO*, is an extension of MIMO, which groups together antennas at the transmitter and receiver on a larger scale to provide better throughput and spectrum efficiency. An example of a massive MIMO is represented by 128-antenna array in a 64-transmit/64-receive configuration. As discussed in [3], massive MIMO can enhance spectral efficiency in two respects: (a) by allowing a base station (BS) to communicate with multiple devices on the same time-frequency-space resources, and (b) by allowing multiple data streams between the BS and each device.

- **Beamforming:** It can be defined as a technology that focuses a wireless signal towards a specific receiving device rather than spreading the signal in all directions. That is, it is a procedure steering the majority of signals generated from an array of transmitting antennas to an intended angular direction [3]. The use of beamforming allows one to deliver high quality signals to the receiver, thus reducing the transfer latency time and the number of errors.

- **Full duplex:** Today BSs and mobile devices rely on transceivers that must take turns when transmitting and receiving information over the same frequency, or operate on different frequencies if a device wishes to transmit and receive information at the same time [4]. With full-duplex technology, a device will be able to transmit and receive data at the same time and on the same frequency. Such technologies can double the capacity of wireless networks.

- **Software defined networks (SDN):** SDNs were introduced to enhance the flexibility of networks by separating network control and forwarding planes and making the control plane and data directly programmable. 5G SDN in combination with network functions virtualization technologies will support the creation of multiple hierarchies that compose the network topology, thus making 5G cellular networks able to meet different application requirements.

## 3. Approaches to the security and privacy analysis of 5G protocols

The brief descriptions above clearly show how the combination of several technologies can greatly increase communication capacity, reduce transmission latency, and save energy. However, it also clearly shows the complex and decentralized nature of 5G cellular networks, which expands cyber risks.

Having a robust 5G ecosystem will require designing protocols (e.g., connecting/disconnecting to/from the network) that are able to achieve their promised security and privacy guarantees even in the presence of adversarial influence. It is thus crucial to analyze designed protocols rigorously against their security and privacy guarantees in the context of an adversarial environment. This is particularly crucial since vulnerabilities in the design are likely to trickle down to implementations/deployments. Analyzing the design-specifications/standards is full of challenges, such as having to reason about stateful protocols that leverage cryptographic constructs. In addition, these complex cellular network protocols can also interact with each other in unanticipated ways lending to the complexity of rigorous and formal protocol analyses. To address such needs, Hussain et al. have developed the first model-based adversarial testing frameworks, LTEInspector [5] and 5GReasoner [6], for systematically analyzing 4G LTE and 5G cellular protocol standards in the context of security and user privacy.

To address the challenge of reasoning about stateful protocols that employ cryptographic constructs, LTEInspector [5] combines the reasoning power of a symbolic model checker (for reasoning on trace properties and temporal ordering of different events/actions) and a cryptographic protocol verifier (for reasoning on cryptographic constructs, e.g., encryption, integrity protections) in the symbolic attacker model using the counter-example guided abstract refinement (CEGAR) principle. In this approach, all cryptography-related details are first abstracted away from the abstract model of the LTE protocol and the desired property to be tested against. The model is then automatically instrumented to include a relaxed Dolev-Yao style adversary who can drop, inject, or sniff messages in the public communication channels without having to adhere to cryptographic assumptions. Model checking is then used to verify whether the property in question holds against the abstract and adversary-instrumented model. If a counter-example demonstrating the property violation is found by the model checker, then a cryptographic protocol verifier is invoked to check whether the counter-example can be realized by an attacker without violating the cryptographic assumptions. This additional refinement step of consulting with a cryptographic protocol verifier is warranted, as the produced counterexample could be a spurious one due to the abstraction of the model. Through its analysis, LTEInspector was able to uncover 10 new attacks in the Non-Access Stratum (NAS) layer procedures of 4G LTE network. This approach of lazily combining the reasoning power of a symbolic model checker and a cryptographic protocol verifier not only enables analysis of cellular protocols but can also be broadly applicable to analyze other real-world complex protocols.

Since the instantiation of LTEInspector framework considers only a single layer (i.e., NAS) of the protocol stack and also does not model packet payload, LTEInspector may miss out on interesting cross-layer and payload-dependent protocol behaviors. To address these limitations, Hussain et al. designed the 5GReasoner [6] framework with a different protocol modeling discipline for 5G networks. Since faithfully capturing all packet payloads impedes the scalability of the analysis, 5GReasoner captures only those packet payloads that impact the security- and privacy-specific behavior of the NAS and RRC (Radio Resource Control) layer protocols. To address the state-explosion problem of the model checking step due to payloads, the framework employs behavior-specific predicate abstractions. With 5GReasoner, Hussain et al. [6] identified eleven new exploitable protocol design weaknesses/flaws.

The initial direction set by the LTEInspector framework also has been followed by other efforts using formal modeling techniques for the analysis of 5G, including the theoretical analysis of the authenticated key exchange protocols used in 5G [7] that use a cryptographic verifier and a fuzzing-based approach to identify design and implementation vulnerabilities in 5G source code by carriers and device vendors [10].

## 4. Defenses

While systematic analysis helps in identifying the root causes of vulnerabilities, it is also paramount to design efficient mitigation techniques and secure solutions to protect the next-generation cellular networks against advanced threats.

**Defense against fake base stations:** A cellular device's connection to the operator's network starts off with the device initiating a connection to the base station that emits signals with the highest strength. Unfortunately, no mechanism currently exists by which a device can verify the legitimacy of a base station in the first place (see also discussion in [15]). This lack of authentication allows adversaries to install rogue base stations, which lure unsuspecting devices to connect to them. Forcing devices to connect to a fake base station is often the necessary first step for the adversary to carry out other destructive attacks, such as man-in-the-middle, location tracking, SMS phishing, relay, and denial-of-service attacks. A notable exploitation of this weakness is performed by typical IMSI catchers, which, after luring the victim device, obtains the victim's permanent identifier IMSI through a benign protocol interaction. Although this fundamental connection bootstrapping weakness is widely acknowledged, there does not seem to be a conscious effort in mitigating this even in the 5G standard that only strives to protect illegitimate exposures of a device's permanent identifier using public-key encryption.

To address this problem of insecure connection bootstrapping, Hussain et al. designed a Public-Key Infrastructure (PKI) based authentication mechanism [9] that leverages precomputation based offline-online digital signature generation algorithms to authenticate the initial broadcast messages emitted by a base station. Offline-online algorithms [18, 19] allow one to split a message to be signed in different portions and pre-compute the signatures of some portions off-line (typically of the portions that are fixed). The signatures of the other portions are computed on-line and then the signatures of all portions are aggregated. The proposed authentication mechanism also employs several optimization schemes to address the requirements of small signature size, efficient signature generation, and short verification time. One of the desired properties of this defense is that it can be deployed incrementally while maintaining backward compatibility.

**Defense against fake emergency alerts:** After the establishment of a secure connection and session keys with the core network and base station, and when a device moves into the battery-conserving idle state after pre-defined interval of inactivity, the core network and base station work in harmony to deliver notification of service (e.g., SMS) to the device using the broadcast paging message. Along with incoming services (e.g., phone call), this unprotected message in both 4G LTE and 5G is also broadcast for sending emergency alerts to the device (e.g., Tsunami, Earthquake, and Amber). Taking this into consideration, Singla et al. have designed PTESLA, a symmetric-key based broadcast authentication mechanism [10] specifically tailored for 4G and 5G networks to protect the devices from unauthorized/fake paging messages.

**Defense against identity exposure attacks:** To prevent illegitimate exposures of permanent identifier (e.g., International Mobile Subscriber Identity or IMSI in 4G, and Subscription Permanent Identifier or SUPI in 5G, respectively) of cellular subscribers, the 3GPP standard [12] for 5G specifications requires a cellular device to encrypt its IMSI/SUPI using the public-key of the core network. Khan et al. [16], however, pointed out that the current 3GPP proposal of IMSI/SUPI protection mechanism still exposes the network operator information of a cellular subscriber and thus cannot meet the strict privacy requirements. To address this limitation, Khan et al. [16] has proposed a 5G-SUPI protection scheme based on Identity based Encryption (IBE), which induces significant protocol overhead.

**Defense against side-channel attacks:** Hussain et al. [8] and Singla et al. [10] have also developed two countermeasures against ToRPEDO style side-channel attacks that exploit the fixed/static paging occasion in 4G and 5G networks [8]. While the solution proposed by Hussain et al. [8], is geared towards backward compatibility and carefully injects and blends noises with the actual paging messages, the solution designed by Singla et al. [10] prescribes variable paging occasion based on the temporary identifier (TMSI).

## 5. Research Roadmap

Even though previous and on-going research efforts represent a good initial step towards security solutions for 5G cellular networks much more work is needed. In what follows, we articulate an initial research roadmap identifying several research directions.

▶ **Formal Analysis of Standards:** Because of the complexity of 5G networks, research activities focusing on formal methods for analyzing the 5G standards encompass different research tasks:

- **Radio Protocol Stack:** While the systematic analysis frameworks developed by Hussain et al. [6] have been successful at identifying design weakness/flaws of the NAS and RRC layer protocols of 4G and 5G cellular networks, violations of security and privacy guarantees in other layers of the protocol stack, e.g., PDCP, RLC, MAC, and PHY (as shown in Figure 1) may lead to attacks and malicious activities, including impersonation attack, website redirection, DDoS, and radio frequency fingerprinting and jamming attacks. It is, therefore, crucial to incorporate such rigorous security and privacy analysis to all layers of 5G protocol, including radio access and core network, and the services that hinge on these critical networks.

- **Inter-networking Protocols:** The inter-networking protocols (e.g., SS7 and Diameter) for connecting multiple network operators have also historically been vulnerable to many different attacks. It is therefore important to analyze these existing inter-networking protocols, identify the trust assumptions across different network operators, and formally and rigorously evaluate them.

- **Network Slicing:** Network slicing, a key feature in 5G networks, will enable several new services, which will potentially expose new classes of security and privacy threats. The reason is that 5G requires each slice to be independent, self-contained, dynamic, and adaptive in order to satisfy the varying needs of the different services. Therefore, it is important to evaluate whether the security and privacy posture meet the trust assumptions and security requirements between tenants and providers.

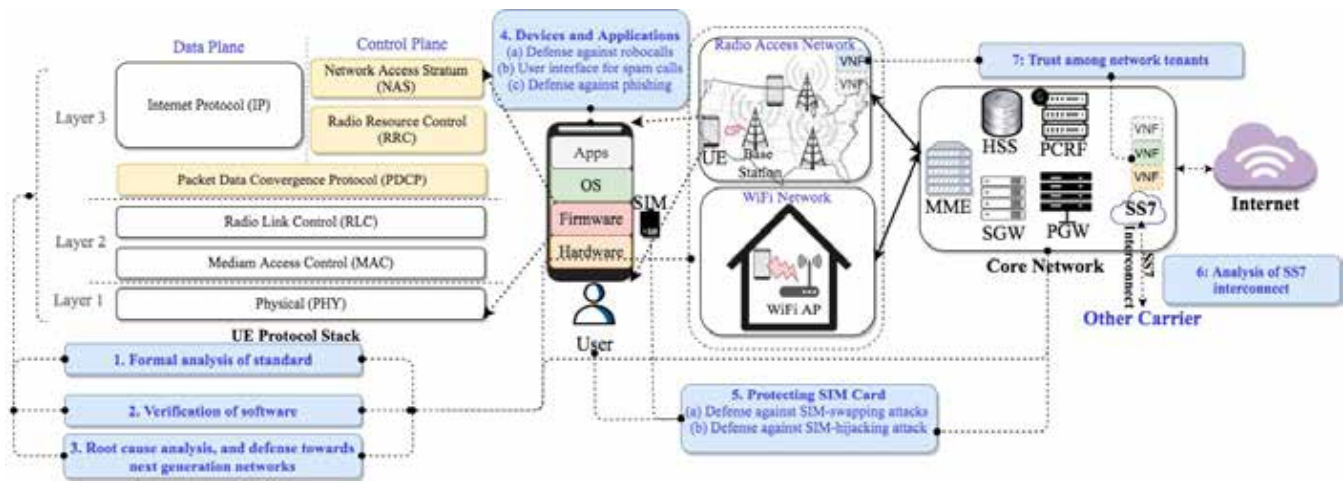

**Figure 1.** Cellular Network Architecture

▸ **Verification of Software and Firmware:**
Implementations often deviate from the design because of specification ambiguities, missing security and privacy requirements, unsafe practices, and oversights stemming from inadequate input sanitization and simplification/optimization of complex protocol interactions [11]. Therefore, it is pivotal to holistically verify whether 5G protocol/system implementations faithfully adhere to the design specifications along with the security and privacy requirements. Also, the various technologies and devices that collectively comprise 5G technology (see Section 2) need to be closely analyzed for security. Depending on the specific technology and devices, firmware may have to be analyzed for security.

▸ **Root Cause Analysis, and Defense Development for Next Generation Cellular Networks:**
Understanding the root cause is important in order to partition protocol-level attacks (including identity exposure, location tracking, denial-of-service, and impersonation attacks) into classes of attacks where attacks in a particular class exploit the same protocol vulnerability. Once such an attack class is identified, one should develop defenses that will thwart that class of attacks by eliminating the underlying protocol vulnerability. To ensure the effectiveness of the attack, one can again analyze the protocol using similar frameworks, such as LTEInspector and 5GReasoner, with the developed defense in place. Note that, band-aid-like defenses, in many cases designed in order to maintain backward compatibility, often do not hold up under detailed scrutiny. For instance, the paging procedure in a 5G network is significantly similar to that of 4G with a minor change in the paging occasion computation, which is now based on the temporary identifier instead of IMSI. However, the recent work on 5G formal verification of NAS and RRC layers has uncovered new vulnerabilities that allow the adversary to break the unlinkability of temporary identifiers and the privacy of paging occasion of a device and thus let the adversary track the user in a target area. Hence, it is pivotal to come up with clean-slate design of important sub-protocols, which may be adopted in future generations of cellular networks.

▸ **Application and Device Security:** Unwanted phone calls, including spam and spoofed robocalls [13] have been a major concern for the last few years. Along with that, many cellular subscribers have recently been hit by the SIM-swapping/hijacking attack [14], which enables the adversary to take over the victim's phone number to cause damage to victim's finances and credit score. It is crucial to first understand the root cause of these attacks and then build verified defenses. To prevent robocalls we need to ensure the integrity of caller identity, whereas to prevent SIM-swapping/hijacking attacks the network operators should enforce two-factor authentication through SMS or phone calls. Overall, one holy-grail to achieve in this domain is the ability to prove end-to-end security and privacy of a given application—that is, composing the application-level security measures and the guarantees provided by the cellular network indeed entail the overall expected security guarantees of an application. Such an approach can be extremely helpful during application development in deciding what security mechanisms an application developer can employ to achieve the application's expected security and privacy guarantees. This will particularly rule out vulnerabilities that stem from the developers' unjustified trust assumptions about the cellular network.

## 6. Concluding Remarks

The 5G ecosystem is complex and involves a large number of parties with different interests and perspectives [20]. It is clear that identifying vulnerabilities and designing defenses cannot be achieved by academic research alone. It requires the collaboration of industries from multiple sectors, including network providers, manufacturers of equipment, software and service companies [17], as well as governmental regulatory and standardization bodies.

**Acknowledgments.** We would like to thank our CCC colleagues Sujata Banerjee, Khari Douglas, Mark Hill, Daniel Lopresti, and Jennifer Rexford for the interesting comments and suggestions on the white paper.

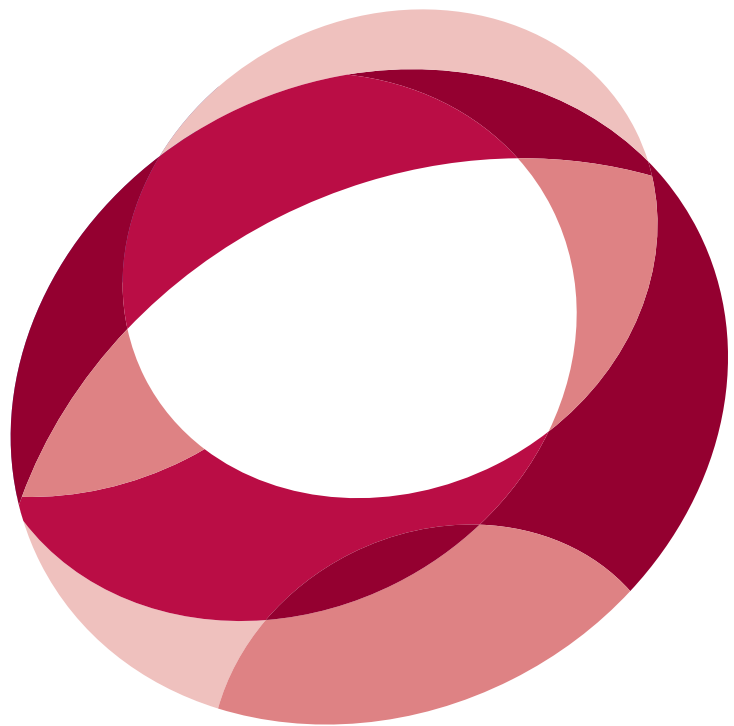